\documentstyle{article}
\title{
\bf  Thin Animals 
}
\author{ {\it D.A. Johnston}\\
         Dept. of Mathematics\\
         Heriot-Watt University\\
         Riccarton\\
         Edinburgh, EH14 4AS, Scotland}

\date {22nd June 1998}         
\begin{document}
  \maketitle
                      {\Large
                      \begin{abstract}
%
Lattice animals provide a discretized model for the $\theta$ transition
displayed by branched polymers in solvent. Exact graph enumeration studies have
given some indications that the phase diagram of such lattice animals
may contain two collapsed phases as well as an extended phase. This has not been confirmed by
studies using other means.
We use the 
exact correspondence between the $q \rightarrow 1$ limit of
an extended Potts model and lattice animals to investigate the
phase diagram of lattice animals on 
$\phi^3$ random graphs of arbitrary topology
(``thin'' random graphs). We find
that only a two
phase structure exists -- there is no sign of a second collapsed phase.

The random graph model is solved in the thermodynamic
limit by saddle point methods. We observe that the ratio of these
saddle point equations give precisely the fixed
points of the recursion relations that 
appear in the solution of the model on the Bethe
lattice by Henkel and Seno. This explains the equality 
of non-universal quantities
such as the critical lines 
for
the Bethe lattice and random graph ensembles. 
%
                        \end{abstract} }
%
  \thispagestyle{empty}
%
%
  \newpage
%
                  \pagenumbering{arabic}

\section{Introduction}

Polymers, whether linear or branched, typically display
a two phase structure in solvent with a
$\theta$-transition \cite{1}, whose position
depends on temperature and solvent
composition, intervening between an extended
and a collapsed phase. 
The case of linear polymers
has proved amenable to various analytical treatments
such as conformal field theory and Coulomb gas methods
\cite{2},
but branched polymers have been rather more stubborn. 

A discretized model of branched polymers is provided by
lattice animals, in essence graphs of connected sites
on some lattice. Contact interactions exist between
nearest neighbour sites that are not directly linked by a bond of
the animal and solvent interactions exist between occupied sites and
unoccupied nearest-neighbour sites. All the ingredients of the
continuum polymer in solvent problem are thus present. It is known that
in $d$ dimensions the lattice animal exponents are related to those
of the Yang-Lee edge singularity in $d-2$ dimensions which allows the
calculation of the bulk entropic exponent for the so-called
strongly embedded lattice animal \cite{3}.
It is also known that the model corresponds exactly to the $q \rightarrow 1$
limit of a certain extended $q$-state Potts model \cite{4,4a}, which has led to
the conjecture that a percolation critical point separates two different
branches of the $\theta$ line \cite{5,5a}. More controversially, exact enumeration
techniques have suggested that {\it two} collapsed phases might exist \cite{6}
whereas transfer matrix investigations have found no sign of this \cite{5a,7}.

This conflict motivated Henkel and Seno \cite{8} to conduct a two-pronged
investigation of lattice animals using the extended
Potts model formalism. They attacked the infinite dimensional, mean-field
problem by solving the model on a Bethe lattice 
\footnote{Although it might appear at first sight that contact interactions
could not be introduced on the loop-less Bethe lattice it should be remembered
that the Potts spins encode the animal state, rather than it being related 
simply to the lattice geometry. If one were really placing an animal on the lattice
directly, some sort of cactus lattice would be required to allow such contacts}
and the 
low-dimensional problem by using Migdal-Kadanoff \cite{4a,9} recursion relations.
In both cases they found only one collapsed phase, and observed that it was
unlikely that the topology of the phase diagram would
change in intermediate dimensions.
In this paper 
we also use the
extended Potts model formalism, this time on
$\phi^3$ random graphs of arbitrary topology, to show that only one
collapsed phase exists for these. 

A couple of interesting technical points emerge along the way:
firstly,
ratios of the saddle point equations in the random graph model
are isomorphic to the fixed points of the recursion relations that solve the
model on the Bethe lattice, so the phase structure on $\phi^3$ graphs
in the large $n$ limit
is identical in all respects to the Bethe lattice; 
secondly, the symmetry breaking pattern that emerges at the $\theta$-transition
is not that which one might have naively expected given earlier results for 
Potts models and percolation on random graphs. 
The former phenomenon
also occurs in other models on random graphs \cite{10,10a,11} and is heuristically explained
by noting that the loops in the random graph model are predominantly large
in the thermodynamic limit \cite{12}, so the random graphs look locally like
the corresponding Bethe lattice and one generically obtains mean field behaviour.

The partition function for lattice animals may be written in several equivalent ways,
but we restrict ourselves to that in equ.(\ref{an1}) to facilitate easy comparison
with \cite{8}.
If an animal contains $n$ sites, $b$ bonds and $k$ contact
interactions we can write 
\begin{equation}
Z_{animal} = \sum_n x^n Z_n ( y, \tau) = \sum_{n,b,k} a_{n,b,k} x^n y^b \tau^k
\label{an1}
\end{equation}
where the site, bond and contact fugacities are given by
$x,y,\tau$ respectively 
and the $a$'s give the number of different animals with the specified numbers of sites, bonds and contacts.
There is an exact correspondence between the lattice animal partition function of equ.(\ref{an1})
and the $q \rightarrow 1$ limit of an extended $q$ state Potts model \cite{4,4a} with hamiltonian
\begin{equation}
{\cal H} = - J \sum_{(i,j)} \delta_{\sigma_i , \sigma_j} - L \sum_{(i,j)} \delta_{\sigma_i , 1} \delta_{\sigma_j, 1} - H \sum_i \delta_{\sigma_i , 1},
\label{ham}
\end{equation}
where the $(i,j)$ sums are over nearest neighbour vertices and the spins $\sigma_i$ take on $q$ values.
The $\sum_{(i,j)} \delta_{\sigma_i , 1} \delta_{\sigma_j, 1}$ term is non-standard, in its absence we would
have the usual Potts hamiltonian (with an external field).

The Potts model parameters are related to $x,y$ and $\tau$ in equ.(\ref{an1}) by
\begin{equation}
x= \exp ( - H - \gamma ( J + L ) ), \; \; y = ( \exp ( J ) - 1) \exp ( J + L ), \; \; \tau = \exp ( J + L)
\end{equation}
where $\gamma$ is the coordination number of the lattice or graph on which the animals live,
and the partition functions are related by
\begin{equation}
Z_{animal} = \lim_{q \rightarrow 1} {\partial \over \partial q} \ln  Z_{Potts}
\end{equation}
where the Potts partition function is
\begin{equation}
Z_{Potts} = 
 \sum_{ \{ \sigma \} } \exp ( - {\cal H} ).
\end{equation}
The idea behind the correspondence is essentially identical to the
use of the standard Potts model Hamiltonian in the $q \rightarrow 1$
limit to describe percolation \cite{13}, which is based
Fortuin and Kastelyn \cite{14} cluster transcription of 
Potts and related models. 

\section{Animals on Thin Graphs - Obtaining the Solution}

Having discussed the lattice animal model in general we now move on to formulate
it on random graphs of arbitrary topology.
A particularly economical way of writing down statistical mechanical models
on such  random graphs was introduced in \cite{15} where it was observed that
taking the $N \rightarrow 1$ limit in the Hermitian matrix models familiar
from studies of 2D gravity generated random graphs of arbitrary topology, as opposed to
the planar graphs picked out by the $N \rightarrow \infty$ limit that is 
used in 2D gravity.
The random graphs appear as Feynman diagrams in the perturbative
expansion of the partition function.
The matrix indices vanish in the $N \rightarrow 1$
limit, leaving ``thin'' graphs rather than the fat, or ribbon, graphs of the
matrix model proper. We shall denote the non-planar, arbitrary topology random graphs as thin graphs
and the animals which live on them as thin animals
throughout the rest of the paper for brevity.

In general the partition function of a statistical mechanical model on an ensemble
of thin random graphs with $2m$ vertices is given by 
\begin{equation}
Z_m \times N_m = {1 \over 2 \pi i} \oint { d \lambda \over
\lambda^{2m + 1}} \int { {\prod_{i} d \phi_i  \over 2 \pi \sqrt{\det K}}
\exp (- S )},
\label{part}
\end{equation} 
where the contour integral over the vertex coupling $\lambda$ picks out
the graphs with $2m$ vertices, $S$ is an appropriate action, $K$
is the inverse of the quadratic form in this action, 
and the $\phi_i$ are the various matter variables in the action.
As the number of thin random graphs increases factorially with $m$
the factor $N_m$ gives the
number of undecorated (i.e. without matter) graphs in the class of interest and disentangles this growth
from any non-analyticity due to phase transitions. For the $\phi^3$ (3-regular) random graphs
we discuss here
\begin{equation}
N_m = \left( {1 \over 6} \right)^{2m} { ( 6 m - 1 ) !! \over ( 2 m ) !!
}.
\end{equation}
It is perhaps worth emphasising that the $\phi_i$ appearing in equ.(\ref{part})
are {\it scalar} variables, so the evaluation of thin graph partition functions
is inherently easier than dealing with the matrix integrals of the planar limit.
In the large $m$, thermodynamic, limit saddle point methods may be used
to evaluate equ.(\ref{part}). The saddle point equation for $\lambda$ may be trivialised
by scaling it out of the action as an overall factor, leaving any critical behaviour
to be revealed by the behaviour of the saddle point equations for the matter fields $\phi_i$.
Phase transitions are manifested by an exchange of dominant saddle point, either continuously
or discontinuously.

We can now write down the action which gives the Boltzmann factors appropriate
to the hamiltonian of equ.(\ref{ham})
for a $q$-state Potts model on thin $\phi^3$ random graphs
\begin{equation}
S = {\alpha \over 2} \psi^2 + {\beta \over 2} \sum_{i=1}^{q-1} \phi_i^2
- \sum_{i=1}^{q-1} \psi \; \phi_i - \delta \sum_{i>j} \phi_i \; \phi_j
- {\nu  \over 3} \psi^3 - \sum_{i=1}^{q-1} { 1 \over 3} \phi_i^3
\label{action}  
\end{equation}
where 
\begin{eqnarray}
\alpha &=& {y \over \tau} + q - 1, \; \; 
\beta = { \tau ( y + ( \tau - 1 ) ( q - 2) ) \over y}, \\ \nonumber
\delta &=& { \tau ( \tau - 1 ) \over  y}, \; \; 
\nu = x^{-1} \tau^{-3},
\end{eqnarray}
and we have pre-emptively scaled out the cubic vertex coupling $\lambda$
as it plays no role in the determination of the phase structure.
The inverse of the quadratic form in equ.(\ref{action})
has the diagonal terms $(\tau, \; (y/ \tau + 1), \; (y/ \tau + 1), \ldots)
= (\exp (J + L ), \; \exp(J), \; \exp(J), \ldots)$
with all off-diagonal terms 1, which is precisely what is required.
We can also see that when $L=0$ we have $\alpha=\beta = \exp(J) + q - 2$, $\delta=1$,
and we recover (up to a trivial rescaling) the action for the standard Potts model
used in \cite{10}.

The saddle point equations for the action in equ.(\ref{action}) may be solved
for increasing $q$ until patience, or algebraic computing power, run out. 
``Generic'' behaviour sets in at $q=4$, as opposed to $q=3$ for the standard action with $L=0$.
One finds a
disordered phase where $\psi$ is distinct and
the $q-1$ $\phi_i$s are equal to $\phi$, which we denote
as having $(1, q-1)$ symmetry.
There are two patterns of symmetry breaking from this phase
to ordered states: in the first we have $\psi$, one distinct $\phi_i = \tilde \phi$ 
and the other $\phi_2 = \phi_3 = \ldots = \phi_{q-1}$ equal to $\phi$, which we denote as $(1,1, q-2)$;
in the second we find $\psi$, $\phi_1 = \phi_2 = \tilde \phi$
and $\phi_3 = \ldots = \phi_{q-1} = \phi$ which we denote by $(1,2,q-3)$. 
We now follow the path used in the analysis
of \cite{10} by writing down effective actions that are sufficient to capture these symmetry breaking patterns.
In effect, most of the $q$ Potts variables are redundant as $q-2$ or $q-3$ of them are equal in the ordered
phases and $q-1$ in the disordered phase.  Imposing the appropriate symmetries we find
the effective actions
\begin{eqnarray}
\label{subdom} 
S_{(1,1,q-2)}  &=& {\alpha \over 2} \psi^2 +{ \beta \over 2 } \tilde \phi^2 + {(q -2) \over 2 } \phi^2 \\ \nonumber
&-&  \psi \tilde \phi - (q - 2) \psi \phi - \delta \left(  \tilde \phi \phi (q - 2) + {1 \over 2} ( q - 2 ) ( q - 3) \phi^2 \right) \\ \nonumber
&-& {\nu  \over 3} \psi^3 - { 1 \over 3} \tilde \phi^3 - { (q - 2) \over 3} \phi^3
\end{eqnarray}
and
\begin{eqnarray}
\label{dom}
S_{(1,2,q-3)} &=& {\alpha \over 2} \psi^2 +\beta \tilde \phi^2 + {(q -3) \over 2 } \phi^2 \\ \nonumber
&-& 2 \psi \tilde \phi - (q - 3) \psi \phi - \delta \left( \tilde \phi^2 + 2 \tilde \phi \phi (q - 3) + {1 \over 2} ( q - 3 ) ( q - 4) \phi^2 \right) \\ \nonumber
&-& {\nu  \over 3} \psi^3 - { 2 \over 3} \tilde \phi^3 - { (q - 3) \over 3} \phi^3.
\end{eqnarray}
Solving the saddle point equations $\partial S / \partial \psi \; = \; \partial S / \partial 
\tilde \phi \; = \; \partial S / \partial \phi \; = \; 0$ for either effective action to get $\psi,\; \tilde \phi, \; \phi$
involves only a quadratic equation in the disordered phase, but one is left with a messy quartic to solve in the 
ordered phase. A direct attack is not very illuminating, but we can use the same legerdemain exercised in \cite{8}
for Bethe lattice to simplify the problem greatly. The useful observation made there was that
the lattice animal partition function could be written as
\begin{equation}
Z_{animal} \sim \sum_n \left( x \exp ( F ) \right)^n
\label{gc}
\end{equation}
where $F$ was the canonical ensemble (fixed $n$) free energy. 
The sum in equ.(\ref{gc}) can be seen to diverge at a critical value
of $x$, which we denote as $\tilde x$,
which in turn gives the critical canonical ensemble
free energy $F = - \ln \tilde x$
of the infinite lattice animal.
Therefore, if we can calculate $ \tilde x$ as a function of $y,\tau$ we obtain
the canonical free energy 
\begin{equation}
F = \lim_{n \rightarrow \infty} n^{-1} \ln Z_n (y, \tau)
\end{equation}
of the infinite lattice animal as a function
of these variables.

To see how this helps, consider the solutions of the saddle point equations obtained
from equs.(\ref{subdom},\ref{dom}) when $q=1$, which are given by
\begin{eqnarray}
\psi = x y \tau^2, \; \; \tilde \Phi^2 - \tilde \Phi + x y = 0, \; \; \Phi^2 - \Phi + x y = 0
\label{solhot}
\end{eqnarray}
for both in the disordered phase, 
with symmetry $(1,q-1)$, where $\tilde \Phi = \tilde \phi / \tau , \; \Phi = \phi / \tau$
and where we have not solved the quadratics explicitly for reasons that will become apparent below.
The ordered phase solutions are obtained from, for equ.(\ref{subdom})  
with breaking pattern $(1,1,q-2)$
\begin{eqnarray}
\label{solsub}
0 &=& \tilde \Phi^4 y^2 + 2 y \left[ 2 ( 1 - \tau) - y \right] \tilde \Phi^3 + \left[ 4 ( \tau - 1)^2 - 6 y ( 1 - \tau) + y^2 ( 1 + x y ) \right] \tilde \Phi^2 \\ \nonumber
&-&  \left[ 4 ( 1 - \tau)^2   + x y^2 ( 2 \tau - y) + 2 y ( \tau - 1) \right] \tilde \Phi + \left[ ( \tau - 1 )^2 + \tau x y^2 \right]  \\ \nonumber
& & \\ \nonumber
\Phi &=&  1 - \tilde \Phi \\ \nonumber
& & \\ \nonumber
\psi &=& - {\tau^2 \over y} \left[ y ( \tilde \Phi^2 - \tilde \Phi ) - 2 ( \tau  - 1) \tilde \Phi + ( \tau - 1) \right] 
\end{eqnarray} 
and, for equ.(\ref{dom}) with breaking pattern $(1,2,q-3)$
\begin{eqnarray} 
\label{soldom}
0 &=& \tilde \Phi^4 y^2 + 2 y \left[ 4 ( \tau - 1 ) - y \right] \tilde \Phi^3 + \left[ 16 ( \tau - 1)^2 + 12 y ( 1 - \tau) + y^2 ( 1 + x y ) \right] \tilde \Phi^2 \\ \nonumber
&-&  \left[ 16 ( 1 - \tau)^2   - x y^2 ( 4 \tau - y) - 4 y ( \tau - 1) \right] \tilde \Phi + \left[ 4 ( \tau - 1 )^2  - 2  \tau x y^2
\right]  \\ \nonumber
& & \\ \nonumber 
\Phi &=&  1 - \tilde \Phi \\ \nonumber
& & \\ \nonumber 
\psi &=& - {\tau^2 \over y} \left[ y ( \tilde \Phi^2 - \tilde \Phi ) + 4 ( \tau  - 1) \tilde \Phi - 2 ( \tau - 1) \right].
\end{eqnarray}
To avoid the onerous task of solving for $\tilde \Phi$ in all of these we now, as advertised,
treat equ.(\ref{solhot}) and
the first of equs.(\ref{solsub},\ref{soldom}) as equations for $x ( y, \tau, \tilde \Phi )$,
in which variable they are only linear.

\section{Animals on Thin Graphs - Discussion of the Solution}
  
If we make the substitution $p = \tilde \Phi + 1/2$ to simplify the resulting expressions
we find from equs.(\ref{solhot},\ref{solsub},\ref{soldom})
\begin{eqnarray}
\label{xs}
x_{1,q-1} (y, p) &=& { 1 - 4 p^2 \over 4  y} \\ \nonumber
x_{1,1,q-2} (y, \tau, p) &=&  { ( 1 - 4 p^2)^2 y^2 + 16 ( \tau - 1) ( 1 - 4 p^2 ) p y + 64 p^2 ( \tau - 1)^2 \over
4  y^2 (( 1 - 4 p^2 ) y  + 8 p \tau) }\\ \nonumber
x_{1,2,q-3} (y, \tau, p) &=& { ( 1 - 4 p^2)^2 y^2 -32  ( \tau - 1) ( 1 - 4 p^2 ) p y + 256 p^2 ( \tau - 1)^2 \over
4 y^2 (( 1 - 4 p^2 ) y  -16  p \tau) }
\end{eqnarray}
where we have labelled the solutions by their symmetry breaking pattern.  
We see here the first hint of a close parallel
with the Bethe lattice solution of \cite{8} since
our two ordered solutions are related by
\begin{equation}
x_{1,2,q-3} (y, \tau, p) = { 1 \over 2} x_{1,1,q-2} ({y \over 2} , \tau, -p),
\end{equation}
which mirrors a similar symmetry in \cite{8}
\footnote{Which does not, however, involve inverting the order parameter.}.

The physical solutions $\tilde x_{1,q-1} (y),\tilde  x_{1,1,q-2} (y, \tau), \tilde  x_{1,2,q-3} (y, \tau)$
are obtained by maximising these expressions with respect to $p$,
which plays the role of an order parameter.
The appropriate solution for a given
$(y, \tau)$ is then the largest of these. 
Bearing in mind that we must have the fugacity $x$ positive to make sense physically
and that the natural ranges of $y$ and $\tau$ are $y \ge 0$, $\tau \ge 1$, we see
that the physical ranges for $p$ are $0 \le p \le 1/2$ for $x_{1,1,q-1}$,  $-1/2 \le p \le 0$ for
$x_{1,2,q-3}$ and $-1/2 \le p \le 1/2$ for $x_{1,q-1}$.

At this point it might appear that we have succeeded in showing precisely
the reverse of what was announced in the introduction in that we have found
{\it two} candidate ordered phases, which might be identified with 
the two contentious collapsed polymer phases. 
However, an extensive numerical investigation throughout
the $(y, \tau)$ range reveals that
the 
one of the solutions, $\tilde x_{1,2,q-3} ( y , \tau)$, is always
larger than the other, $\tilde x_{1,1,q-2} ( y , \tau)$, 
for a given $(y, \tau)$ so it is
the former which is always dominant. 

For the disordered solution we 
clearly have a maximum at $p=0$
\begin{equation}
\tilde x_{1,q-1} (y) = { 1 \over 4 y },
\end{equation}
so the free energy is $F = \ln ( 4 y)$ and we also find
\begin{equation}
x_{1,1,q-2} (y, \tau, 0) =  x_{1,2,q-3} (y, \tau, 0) = { 1 \over 4 y }.
\end{equation}  
We are now in a position to map out the phase structure of thin animals.
We have found two phases, whose
free energies are given by $F = - \ln \tilde x_{1,q-1} (y)$ which we
can identify with the extended phase and $F = - \ln \tilde x_{1,2,q-3} (y, \tau)$
which we identify with the (single) collapsed phase.

The transition line between these two phases may be pinned down by noting that
\begin{equation}
{\partial x_{1,2,q-3} (y, \tau, p) \over \partial p} = 0
\end{equation}
iff $\tau=2$.
We then observe that
\begin{equation}
{\partial^2 x_{1,2,q-3} (y, 2, 0) \over \partial p^2} = 2 { 64 - y^2 \over y^3}
\end{equation}
so $\tilde x_{1,2,q-3} (y , 2, p)$ has a maximum at $p=0$ if $y>8$, a minimum if $y<8$
and a turning point if $y=8$. A glance at Figs.1,2 where we plot 
$x_{1,q-1} (y, p)$ and $x_{1,2,q-3} (y, 2, p)$ for $y=10$ and $y=4$ respectively
clarifies what is going on. In Fig.1 we see that 
$x_{1,q-1} (y, p)$ and $x_{1,2,q-3} (y, 2, p)$  share a common
maximum at $p=0$ when $y=10$ (which is the case for all $y \ge 8$), whereas
for $y=4$ the maximum of $x_{1,2,q-3} (y, 2, p)$ lies at some $p < 0$.
This means that we see a continuous transition as $\tau$ is varied across $\tau=2$
when $y > 8$ between extended animals at small $\tau$ and collapsed animals at large $\tau$.
For $y<8$ the position of the critical line in the $(y, \tau)$ plane must be determined
numerically and the transition is first order. The physical solution jumps 
discontinuously from the maximum
of $x_{1,q-1}$ at $p=0$  to the maximum of $x_{1,2,q-3}$ at $p < 0$. 
The appropriate equations for pinpointing the first order line are thus
\begin{eqnarray}
x_{1,2,q-3} (y, \tau, p) = x_{1,q-1} (y, 0) &=& { 1 \over 4 y} \\ \nonumber
{\partial x_{1,2,q-3} (y, \tau, p) \over \partial p} &=& 0 \\ \nonumber
{\partial^2 x_{1,2,q-3} (y, \tau, p) \over \partial p^2} &\le& 0
\end{eqnarray}
which give the conditions for the maximum in $x_{1,2,q-3} (y, \tau, p)$
at $p < 0$ to be equal to the maximum in $x_{1,q-1} (y,p)$ at $p=0$.

A triple point at $y=8,\tau=2$ separates
the first and second order transition lines. All of these features are indicated in
the schematic phase diagram of Fig.3. 
The presence of a jump across the 
first order line is confirmed explicitly 
by the sample values in Figs.4,5 where we see that as we cross the line
by varying $\tau$ at fixed $y=4$ we move from $x_{1,q-1}$ dominant in Fig.4 to the situation
in Fig.5
where $x_{1,2,q-3}$ is just equal to the maximum of $x_{1,q-1}$ ($1/4y$ at the origin)
for a {\it non-zero} $p$ value ($\simeq -0.1634\ldots$) at $\tau \simeq 1.8528\ldots$.
At this point $p$ jumps discontinuously and we move into the collapsed phase as $\tau$
increases.

When moving across the second order portion of the line however,
the maximum of $x_{1,2,q-3}$ is at the origin and we see no jump in $p$, giving
a second order transition. This can be seen by referring back to Fig.1 and
comparing it with Fig.6. In Fig.6 we are {\it below} the second order line
at $y=10, \tau=1$ and the two curves look much like Fig.4. As we move up
to the line at $y=10, \tau=2$ in Fig.1 we can see that the maximum
of $x_{1,2,q-3}$ now lies at the origin and increasing $\tau$ further
takes us into the collapsed phase via a second order transition.
                   
In closing this section it should be remarked that 
although the broad features of the random graph phase diagram appear
to be in accordance with two-dimensional models, there is one notable
difference. The triple point separating the first and second order portions of the $\theta$ line
is the percolation point in $2D$. That is {\it not} the case here -- the percolation point in
the random graph model is at $L=H=0$ and $\tau=2$, which in turn implies $y = \tau ( \tau - 1) = 2$
\footnote{Remarkably, the same numerical values as the $2D$ model.}.
This presumably can be ascribed to the mean-field nature of the model, a comment which also applies to the Bethe
lattice results of \cite{8}, whose similarities with the work here we now move on to discuss.

\section{Comparison with Bethe Lattice results}

Anyone familiar with the results of \cite{8} for Bethe lattice animals
may at this point be experiencing a strong sense of d\'ej\`a vu.
The thin animal phase diagram has proved to be identical to that on the 
Bethe lattice, down to non-universal quantities such as the position
of the $\theta$ line and triple point. As we have already noted,
even before extracting the physical
solutions $\tilde x$ parallels are apparent since
the two ordered solutions on both random graphs
and the Bethe lattice possess a particular $y \rightarrow y/2$ symmetry. 
Our criteria for extracting the $\theta$ line are also identical to
those used on the Bethe lattice, 
although the equations for $x$ on the Bethe lattice are quadratic
rather than linear, as above.
All this clearly begs an explanation.

Our saddle point equations for $x_{1,2,q-3}$ at $q=1$, which we have so far 
not written explicitly, are given by
\begin{eqnarray}
\label{ours}
\psi^2 &=& {x y \tau^2 } \psi - 2 x \tau^3 \tilde \phi + 2 x \tau^3 \phi  \\ \nonumber
\tilde \phi^2  &=& -  \psi  - { 2\tau ( \tau -1) -\tau y \over y} \tilde \phi + {2 \tau ( \tau  - 1 ) \over y}
\phi  \\ \nonumber
\phi^2 &=& - \psi - {2 \tau ( \tau -1 ) \over y } \tilde \phi + { 2 \tau ( \tau -1) +\tau y \over y} \phi.
\end{eqnarray}
At first sight these bear little relation to the fixed points
of the recursion relations used in \cite{8}
to obtain the dominant solution to the model on the Bethe lattice.

The method of derivation is certainly completely different.
On the Bethe lattice one considers the behaviour of the central spin
$\sigma_0$ \cite{8}, writing
\begin{equation}
Z_{Potts} = \sum_{\sigma_0} \exp \left( H \delta_{\sigma_0, 1} \right) 
\sum_{ \{ s \} } \; \prod_{j=1}^3 \; Q_n ( \sigma_0 | s^{(j)} )
\end{equation}
where $s^{(j)}$ is the spin on the $j$th sub-branch and
\begin{equation}
Q_n ( \sigma_0 | s^{(j)} )  = 
\exp ( J \delta_{\sigma_0, s_1} + L \delta_{\sigma_0,1} \delta_{s_1 ,1} + H \delta_{\sigma_0 , 1} )
\; 
\exp \left( J {\sum_{(i,j)}}' \delta_{s_i,s_j} + L {\sum_{(i,j)}}' \delta_{s_i,1} \delta_{s_j,1} + H {\sum_i}' \delta_{s_i,1} \right) 
\end{equation}
and the primed sums are over the sub-branch with first spin $s_1$.
The solution then follows by defining
\begin{equation}
g_n ( \sigma_0 ) = {\sum_{ \{ s \} }}' Q_n ( \sigma_0 | s ),
\end{equation}
where $n$ labels the ``shell'' of the Bethe lattice,
and noting that it satisfies
the recursion relation
\begin{equation}
g_n ( \sigma_0 ) = \sum_{s_1=1}^q \exp \left( J \delta_{\sigma_0,s_1} + L \delta_{\sigma_0,1} \delta_{s_1,1} + H \delta_{s_1,1} \right)
\left( g_{n-1} (s_1) \right)^2.
\label{curse}
\end{equation}
Defining
\begin{equation}
\label{hisrat}
\Theta_n = { g_n(\sigma_0 \ne 1 , 2 ) \over g_n ( 1)}, \; \; Z_n = { g_n(2) \over g_n(1)}
\end{equation}
and taking the limit $n \rightarrow \infty$ allows one to write the fixed point of the recursion relations
in equ.(\ref{curse})
as
\begin{eqnarray}
\label{his}
\Theta &=& { x^{-1} \tau^{-3} + 2 Z^2 + ( { y \over \tau} - 2 ) \Theta^2
\over x^{-1} \tau^{-2} + 2 Z^2 - 2 \Theta^2 } \\ \nonumber
Z &=& { x^{-1} \tau^{-3} + ( { y \over \tau} + 2 ) Z^2 - 2 \Theta^2 
\over x^{-1} \tau^{-2} + 2 Z^2 - 2 \Theta^2 }
\end{eqnarray}
where $\Theta = \lim_{n \rightarrow \infty} \Theta_n$, $Z = \lim_{n \rightarrow \infty} Z_n$. 

Given the identical phase diagram we have found on random graphs
to the Bethe lattice results of \cite{8} it is clear that the
two sets of equations , equs.(\ref{ours}, \ref{his}), must somehow
be equivalent. This can be made explicit
by a simple transformation. Equs.(\ref{ours})
may be symbolically  written as $\vec \phi^2 = A \vec \phi $,
where $\vec \phi$ is the column vector $(\psi, \tilde \phi, \phi)$
and  $\vec \phi^2$ is $(\psi^2, \tilde \phi^2, \phi^2)$.
We are at liberty to rewrite these as $\vec \phi = A^{-1} \vec \phi^2$
as $A$ is invertible in general 
\begin{eqnarray}
\label{ours2}
\psi &=& {1 \over \tau^2 x y } \psi^2 + { 2 \over y} \tilde \phi^2 - { 2 \over y} \phi^2  \; \; \; \; \; \; \; (a) \\ \nonumber
\tilde \phi &=& { 1 \over \tau^3 x y } \psi^2 + { y + 2 \tau \over \tau y} \tilde \phi^2 - { 2 \over y} \phi^2 
\; \; \; \; \; (b)  \\ \nonumber
\phi &=& { 1 \over \tau^3 x y } \psi^2 + { 2 \over y} \tilde \phi^2 + { y - 2 \tau \over \tau y} \phi^2 \; \; \; \; \; (c).
\end{eqnarray}
If we now divide $(c) /  (a)$ and $(b) / (a)$ 
\begin{eqnarray}
{ \phi \over \psi} &=& { x^{-1} \tau^{-3} \psi^2 + { 2 } \tilde \phi^2 + (  { y \over \tau} - 2 ) \phi^2 \over
x^{-1} \tau^{-2} \psi^2 + { 2 } \tilde \phi^2 - { 2 } \phi^2} \\ \nonumber
{\tilde \phi\over \psi}  &=& { x^{-1} \tau^{-3} \psi^2 + ( { y \over \tau} + 2 ) \tilde \phi^2 - { 2 } \phi^2
\over x^{-1} \tau^{-2} \psi^2 + { 2 } \tilde \phi^2 - { 2 } \phi^2}
\end{eqnarray}
and make the identifications
\begin{equation}
\label{trans}
\Theta = {\phi \over \psi}, \; \; Z = { \tilde \phi \over \psi}
\end{equation}
we obtain exactly the two recursion equation fixed points of equ.(\ref{his}).

Although we have two distinct physical systems, closed random graphs
of arbitrary topology which {\it do} contain loops, and the loop-less Bethe lattice
the preceding calculation shows that the thermodynamic behaviour of lattice
animals on both is identical. In a precise sense the loops do not matter, as the saddle
point equations that govern the random graph thermodynamic limit are isomorphic
to the recursion relations that arise in solving the loop-less Bethe lattice.
The identity between the Bethe lattice recursion relations and ratios of the random graph
saddle point equations also occurs for other systems (standard Ising, Potts spins etc.)
\cite{11}, where the consequences are the same, namely the random graph solutions are 
identical to those on the Bethe lattice.

\section{Discussion}

The astute reader might remark that the solution of the lattice animal problem on random
graphs presented in this paper
does not really provide any further evidence against the existence of
a second collapsed phase, as we have also shown that the solutions to the model
have the same content as the Bethe lattice solution in \cite{8}, though
this is an interesting result in itself.
One could counter the objection by stating that we have found a different physical system,
and solved it by very different means, in order to arrive at these conclusions.

The thin graph solution also clarifies one lacuna in the Bethe lattice approach.
As we have noted, solving the full set of saddle point equations
$\partial S /  \partial \psi = \partial S / \partial \phi_i = 0$ (i.e.
{\it before} going to the effective actions) shows that only one
disordered and two ordered phases
\footnote{One of which is always dominant.} are present for arbitrary $q$.
On the Bethe lattice enough
order parameters are introduced by hand to distinguish the two putative ordered
phases, which is then shown to be self-consistent. The possibility of a more
exotic phase diagram is not, however, explicitly excluded.
The random graph saddle point equations show that this really is enough,
as no other patterns of symmetry breaking are observed for any $q$.

The observed symmetry breaking pattern itself on random graphs is 
unexpected, namely $(1,q-1) \rightarrow (1,2,q-3)$. For the
standard $q$-state Potts model one has $(q) \rightarrow (1,q-1)$,
so the pattern observed for the sub-dominant solution $(1,q-1) \rightarrow
(1,1,q-2)$ might have been a more obvious first guess.
We have also remarked that the critical percolation point,
$y = \tau=2$ on thin graphs, is not equal to the
tricritical point $y= 8, \tau =2$ that separates the two portions of $\theta$ line,
in contrast to the $2D$ model, although the general
topology of the phase diagram is similar.

Although the work presented in this paper is fairly self-contained one 
obvious generalisation suggests itself. The action of equ.(\ref{action})
will also serve to represent lattice animals on {\it planar} random graphs
if one takes the $\psi, \phi_i$ to be $N \times N$ Hermitian matrices and performs
the limit $N \rightarrow \infty$. The $L=H=0$ case (percolation) has already been solved
by Kazakov \cite{kaz}, which suggest that the animal problem might also be tractable.
The interest in doing this is that it could shed light on a another vexed question
for animals -- the (lack of) conformal invariance in the 2D model.

As a technical point, sidestepping the solution of difficult saddle
point equations by considering the canonical free energy and solving
for the fugacity associated with the external field is an idea
which might also be fruitfully applied to the solution of the standard Potts
(or even Ising) models in an external field on random graphs, where
similarly complicated equations arise.

I would like to thank Malte Henkel for helpful discussions,
both in person and electronically.
 
\clearpage \newpage

\clearpage \newpage
\begin{figure}[htb]
\vskip 20.0truecm
\includegraphics{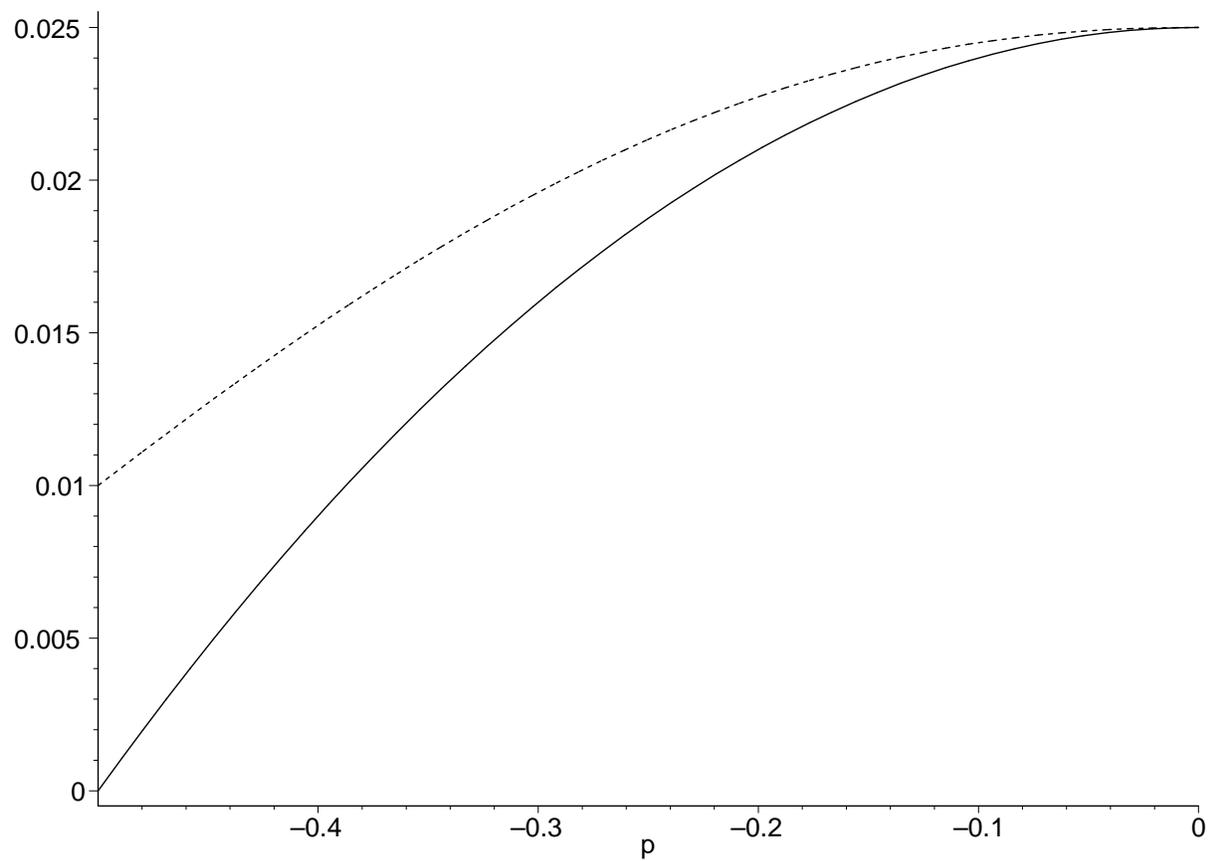}
\caption[]{\label{fig1} The two solutions vs $p$ when $\tau=2, y=10$.
The upper dotted curve is $x_{1,2,q-3}$ and the lower 
bold curve is $x_{1,q-1}$.
The maximum of both curves lies at $p=0$.}
\end{figure}
\clearpage \newpage
\begin{figure}[htb]
\vskip 20.0truecm
\includegraphics{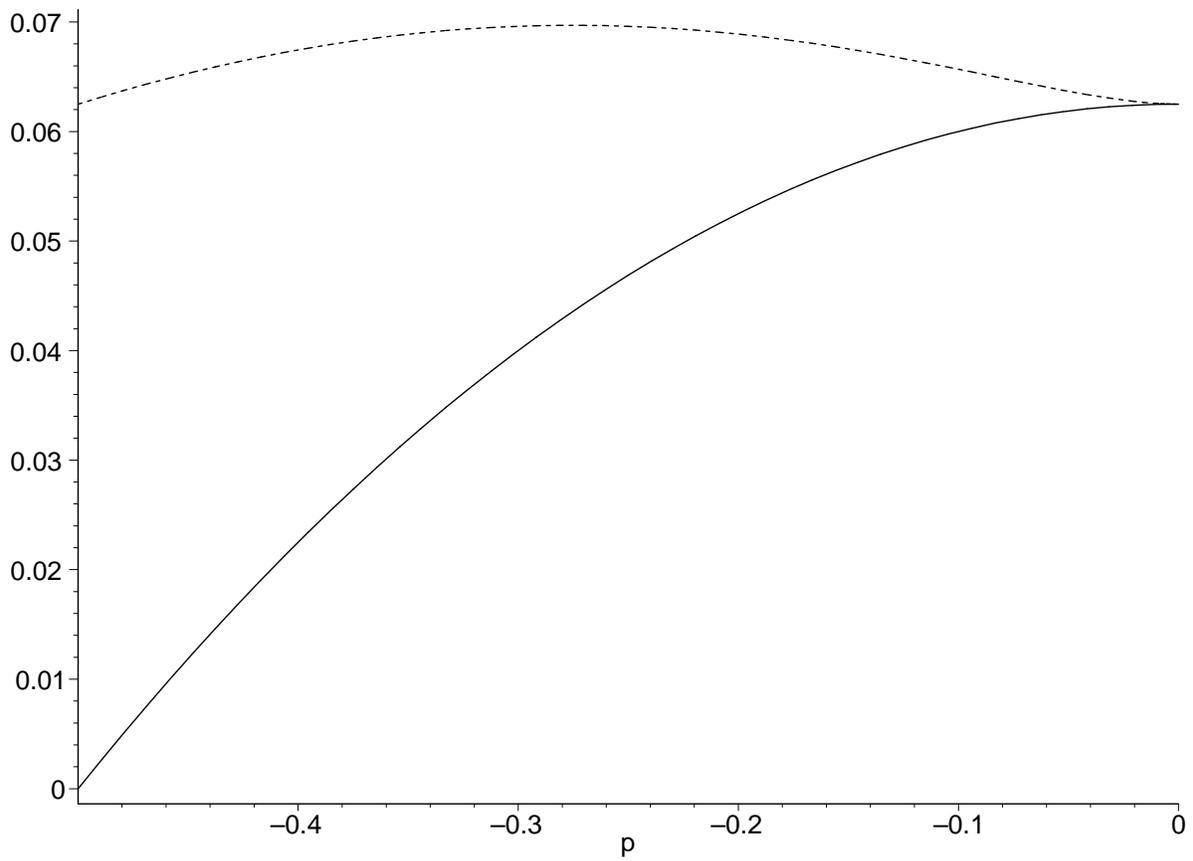}
\caption[]{\label{fig2} The two solutions vs $p$ when $\tau=2, y=4$.
Again, the upper dotted curve is $x_{1,2,q-3}$ and the lower bold curve $x_{1,q-1}$.
The maximum of $x_{1,2,q-3}$ now lies away from $p=0$, signalling
 a first order transition.}
\end{figure}
\clearpage \newpage
\begin{figure}[htb]
\vskip 20.0truecm
\includegraphics{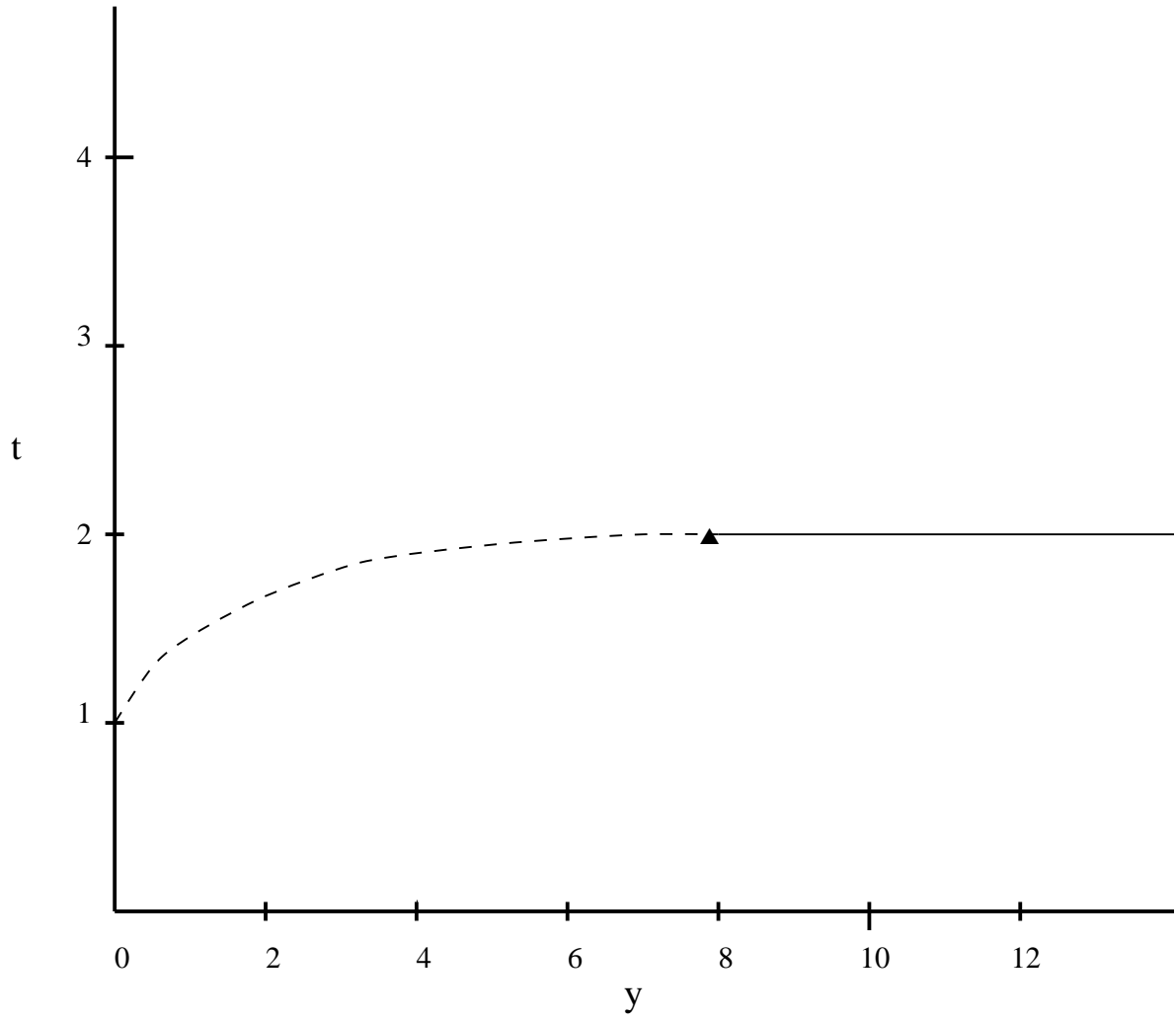}
\caption[]{\label{fig3} The (schematic) phase diagram for thin animals.
The solid line $\tau=2, y > 8$ is a line of second order transitions,
whereas the dashed line is first order. The triple point is
shown at $y=8, \tau=2$ at the meeting point of the two.}
\end{figure}
\begin{figure}[htb]
\vskip 20.0truecm
\includegraphics{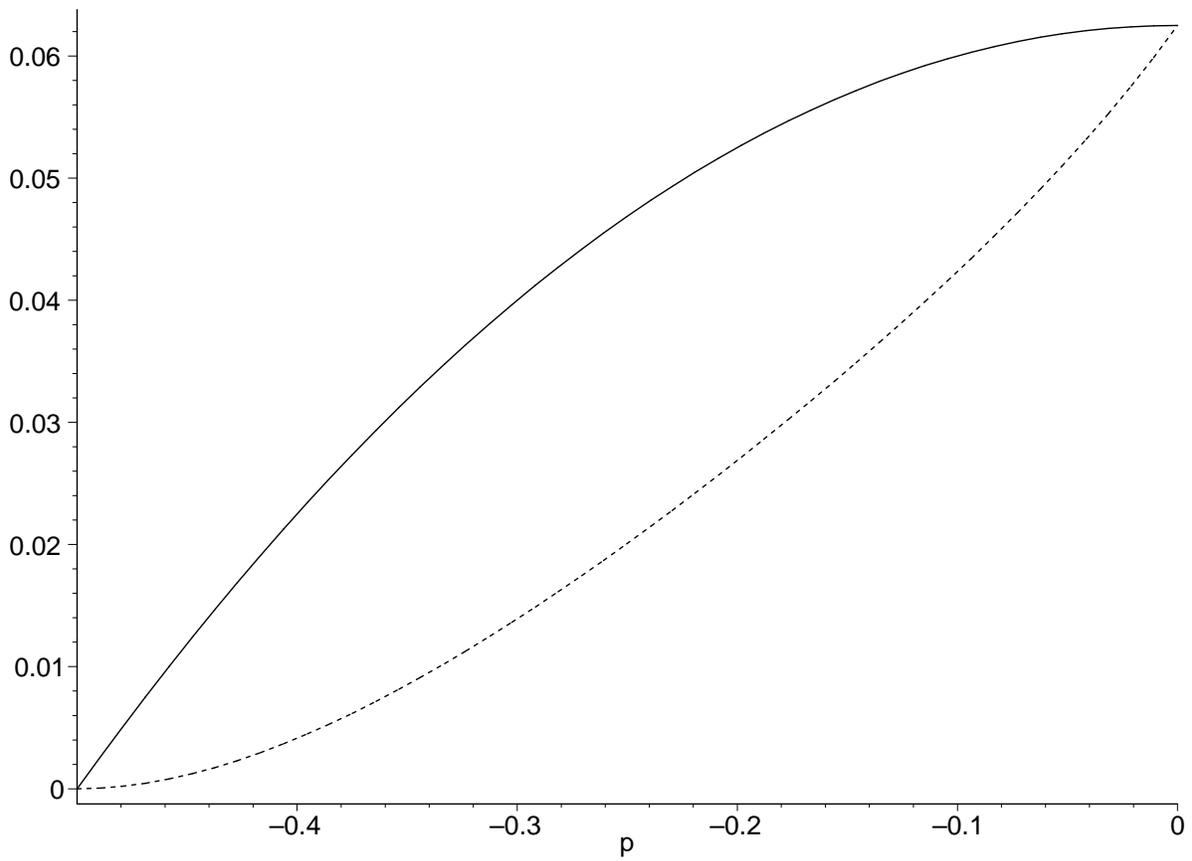}
\caption[]{\label{fig4} The two solutions {\it below} the first order
portion of the critical line at $y=4, \tau=1$. $x_{1,q-1}$ is
shown in bold, $x_{1,2,q-3}$ dotted.}
\end{figure}
\begin{figure}[htb] 
\vskip 20.0truecm 
\includegraphics{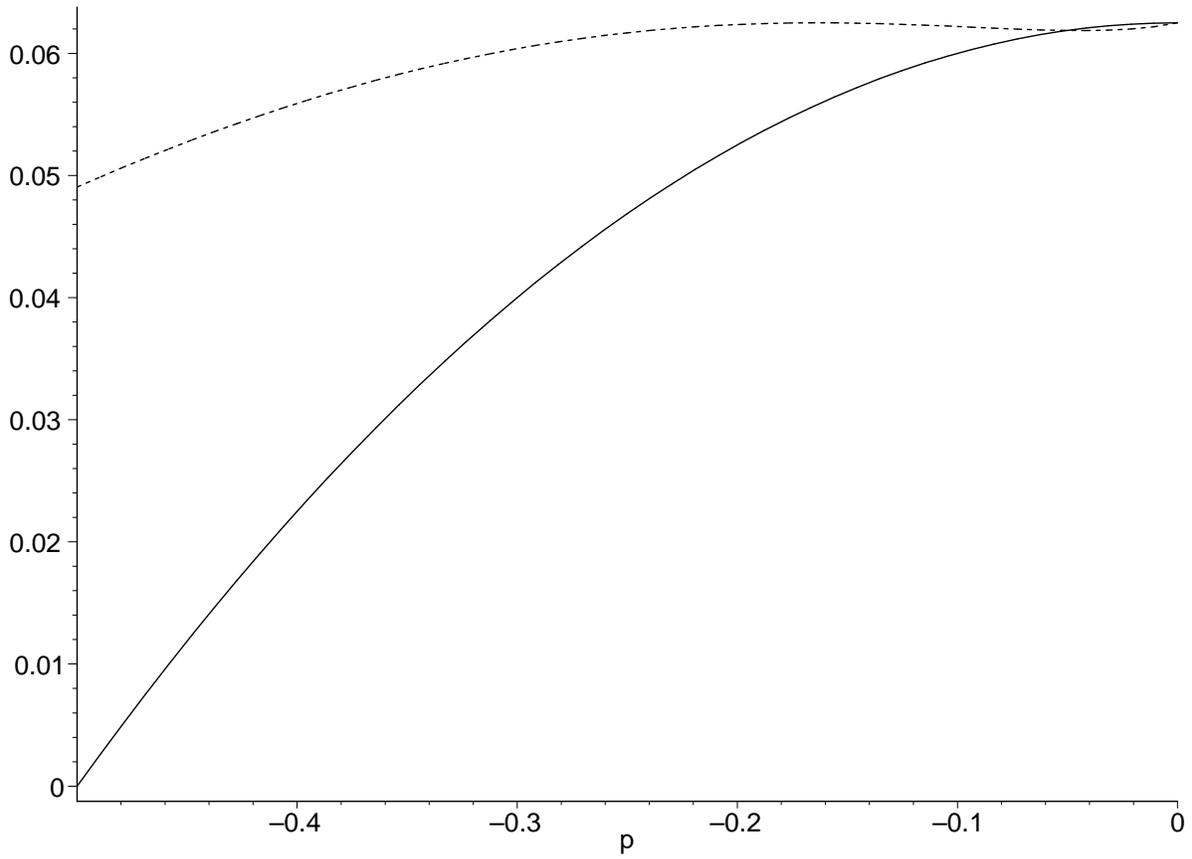} 
\caption[]{\label{fig5} The solutions on the first order
portion of the critical line at $y=4, \tau \simeq 1.8528\ldots$.
$x_{1,2,q-3}$ is shown dotted and
its maximum at $p\simeq -0.1634\ldots$ has
just reached $1/4y$.
$x_{1,q-1}$, shown in bold, has the same maximum at $p=0$.} 
\end{figure} 
\begin{figure}[htb]  
\vskip 20.0truecm  
\includegraphics{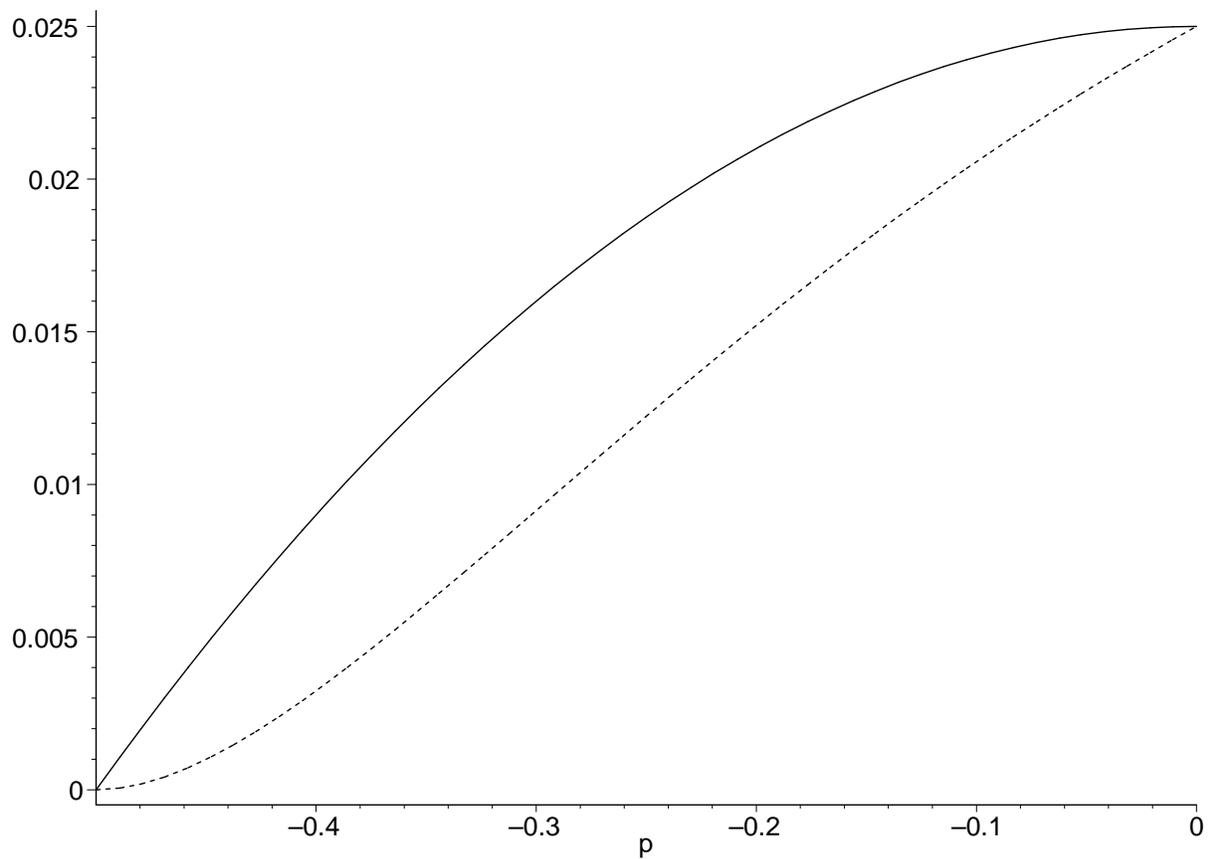}  
\caption[]{\label{fig6} The two solutions {\it below} the second order
portion of the critical line at $y=10, \tau=1$.  
$x_{1,q-1}$, again denoted by a solid line, is dominant.} 
\end{figure}  

\end{document}